\documentclass[journal]{IEEEtran}
\usepackage[utf8]{inputenc} 
\usepackage{cite}
\usepackage{amsmath,amssymb,amsfonts}
\usepackage[linesnumbered,ruled]{algorithm2e}
\usepackage{graphicx}
\usepackage{textcomp}
\usepackage{xcolor}
\usepackage{epsfig}
\usepackage{multirow}
\usepackage{dblfloatfix}
\usepackage{subfigure}
\usepackage{caption}

\begin{document}
\title{A Real Time 1280x720 Object Detection Chip With 585MB/s Memory Traffic}

\author{Kuo-Wei~Chang, ~Hsu-Tung Shih,
	Tian-Sheuan~Chang,~\IEEEmembership{Senior Member,~IEEE}, ~Shang-Hong Tsai,\\ ~Chih-Chyau Yang, ~Chien-Ming Wu,~\IEEEmembership{Member,~IEEE}, ~and Chun-Ming Huang,~\IEEEmembership{Member,~IEEE} %
\thanks{This work was supported by the Ministry of Science and Technology, Taiwan, under Grant 109-2634-F-009 -022,  109-2639-E-009-001 and 110-2622-8-009-018-SB, and TSMC. K. Chang, H. Shih, and T. Chang are with Institute of Electronics, National Yang Ming Chiao Tung University, Taiwan. (e-mail: tschang@nycu.edu.tw).	T. Chang, S. Tsai, C. Yang, C. Wu, and C. Huang are with Taiwan Semiconductor Research Institute, Hsinchu, Taiwan.}
\thanks{
© 2021 IEEE.  Personal use of this material is permitted.  Permission from IEEE must be obtained for all other uses, in any current or future media, including reprinting/republishing this material for advertising or promotional purposes, creating new collective works, for resale or redistribution to servers or lists, or reuse of any copyrighted component of this work in other works.\\
K. -W. Chang et al., "A Real-Time 1280,x,720 Object Detection Chip With 585 MB/s Memory Traffic," in IEEE Transactions on Very Large Scale Integration (VLSI) Systems, doi: 10.1109/TVLSI.2022.3149768.
}
\thanks{Manuscript received Sep. 8, 2021; revised Dec. 5, 2021, and Jan. 5, 2022.}}
\maketitle

\begin{abstract}
Memory bandwidth has become the real-time bottleneck of current deep learning accelerators (DLA), particularly for high definition (HD) object detection. Under resource constraints, this paper proposes a low memory traffic DLA chip with joint hardware and software optimization. To maximize hardware utilization under memory bandwidth, we morph and fuse the object detection model into a group fusion-ready model to reduce intermediate data access. This reduces the YOLOv2's feature memory traffic from 2.9 GB/s to 0.15 GB/s. To support group fusion, our previous DLA based hardware employes a unified buffer with write-masking for simple layer-by-layer processing in a fusion group. When compared to our previous DLA with the same PE numbers, the chip implemented in a TSMC 40nm process supports 1280x720@30FPS object detection and consumes 7.9X less external DRAM access energy, from 2607 mJ to 327.6 mJ.

\end{abstract}


%
\IEEEpeerreviewmaketitle

\section{Introduction}

\IEEEPARstart{O}{bject} detection with deep learning has attracted significant research attention in recent years due to its wide success over traditional computer vision methods\cite{YOLOv2, jiao2019survey}. However, with deeper and wider deep learning models and larger input size, real-time model execution poses challenges of high computation cost and memory bandwidth, especially for edge devices as in autonomous driving. Thus, hardware acceleration with deep learning accelerators (DLAs) is required to tackle these challenges. 

Many DLAs have been proposed with massive parallel processing elements (PEs) to solve the high computation cost and different data reuse policies to reduce memory traffic. We can classify these designs into layer-by-layer processing or layer fusion processing according to their hardware scheduling. The widely used layer-by-layer DLAs process one layer after another, which only needs to store one layer of weights and partial feature maps inside a chip. They reduce memory traffic by different data reuse\cite{eyeriss,dna, VWA}, larger on-chip DRAM macro \cite{Diannao1,Diannao2,Diannao3,Diannao4,Diannao5}, precision adaptive design \cite{envision,thinker_2018jssc,unpu_2018jssc}, and sparse convolution \cite{eyerissv2,vscnn,Hsiao.et.al._JETCAS, STICKER_2020JSSC}. 
In which, major external DRAM bandwidth savings come from reusing input, weights, or partial sums of output \cite{eyeriss}. Similar approaches have also been adopted in DLAs specific to object detection\cite{Nguyen2019, wang2020sparse, han2020design}. In addition, most of those object detection designs use abundant memory resources in FPGA \cite{Nguyen2019} to avoid frequent external DRAM access, which does not apply to ASIC designs for edge devices.
These data reuse policies are limited to one-layer processing only. All these layer-by-layer DLAs have to save per layer output to the external DRAM and load it back for next layer processing, which causes high memory traffic and inhibits further possible processing speedup due to lack of data. 


In contrast, layer-fusion DLAs \cite{fused_layer, MTK_fusion_isscc, SRNPU_JETCAS} fuse multiple-layer computation that processes the next layer once its input data is ready. Thus, it only needs to load the input data of the first layer and save the output data of the last layer in the fusion group and can save the access of intermediate data. 
The fusion requires a large buffer to store intermediate feature maps for fusion processing. To avoid this, the input is usually partitioned into tiles for processing. However, the overlapped areas between tiles have to be stored and recomputed due to data dependency, which results in significant buffer cost when more layers are fused. Thus, non-overlapped tile processing proposed by split-CNN\cite{jin2019split} or block convolution\cite{blockconv} avoids this without considering the overlapped area and still achieves similar accuracy as the original model.  Previous non-overlapped tile processing is targeted on GPU or FPGA, which assumes a large enough buffer for weight storage. However, DLA with ASIC used in edge devices has a much smaller buffer. Note that the layer fusion assumes that all weights within the fusion group should be stored on the chip to avoid repeatedly access during tile processing. This poses design difficulties for a modern network, whose per layer weight number could be easily over 1 M, exceeds buffer size and thus inhibits the possibility of layer fusion. Such a problem cannot be solved by hardware design only as in previous approaches, which needs model adaption for better algorithm and architecture co-design.  



This work offers a fusion ready hardware and software with collaborative optimization to develop a low memory traffic DLA for HD size object detection to address the aforementioned difficulties. The baseline model, YOLO-v2\cite{YOLOv2}, needs 55.6M parameters with external memory traffic of 1.6GB weight and 4.6GB feature map for 1280x720@30FPS execution. 
We propose the resource-constrained network fusion and pruning (RCNet) to make the model group fusion ready.
The preceding process has also combined hardware-specific guidelines to make the model hardware friendly. This allows the model to fit within the limits of weight and feature buffer size, maximizing the benefits of layer fusion and hardware utilization. Following model optimization, the hardware based on our prior design \cite{VWA} uses the unified buffer design with write masking to support group fusion. Its computing flow also employs the modified nonoverlapped tile processing \cite{jin2019split, blockconv} to reduce the data dependency between tiles. 
The final chip is implemented with the TSMC 40nm process, which achieves real-time high definition object detection and needs 7.9X lower external DRAM energy than our previous design  \cite{VWA} with the same PE numbers.

The rest of the paper is organized as follows. Section II introduces the proposed resource-constrained network fusion and pruning for YOLO-v2. Section III presents the proposed architecture. The implementation results and comparisons are shown in Section IV. Finally, this paper is concluded in Section V.

\section{Resource-Constrained Network Fusion and Pruning (RCNet) for YOLO-v2}

\subsection{Overview}
The main limits for maximizing the benefits of layer fusion are the weight and feature map buffer sizes. If the weight buffer cannot hold all weights in a fusion group, weights must be read repeately from external memory for each tile processing, resulting high memory traffic.  The tile size is determined by the size of the feature map buffer, which is a trade-off between accuracy and hardware cost.

For object detection, we choose YOLO-v2 as our baseline model for its simplicity. YOLO-v2 needs 55.6M parameters, which makes layer fusionvery unfeasible since each layer parameter could easily exceed the weight buffer capacity of edge devices. To make it fusion ready while maximize hardware utilization, we propose using RCNet to morph the model according to the buffer size. This resource constrained YOLO-v2 is denoted as RC-YOLOv2.


\subsection{Lightweight model conversion}
The backbone of the original YOLO-v2 is a simple stack CNN structure like VGG\cite{VGG}. This model is not ready for layer fusion due to its large model size. For a fusion-ready model, the weight numbers of any two consecutive layers shall be smaller than the weight buffer size after RCNet. Otherwise, the fusion operation will be degenerated to a layer-by-layer one.

\begin{figure}[t]
	\centering{\includegraphics[width=0.45\textwidth]{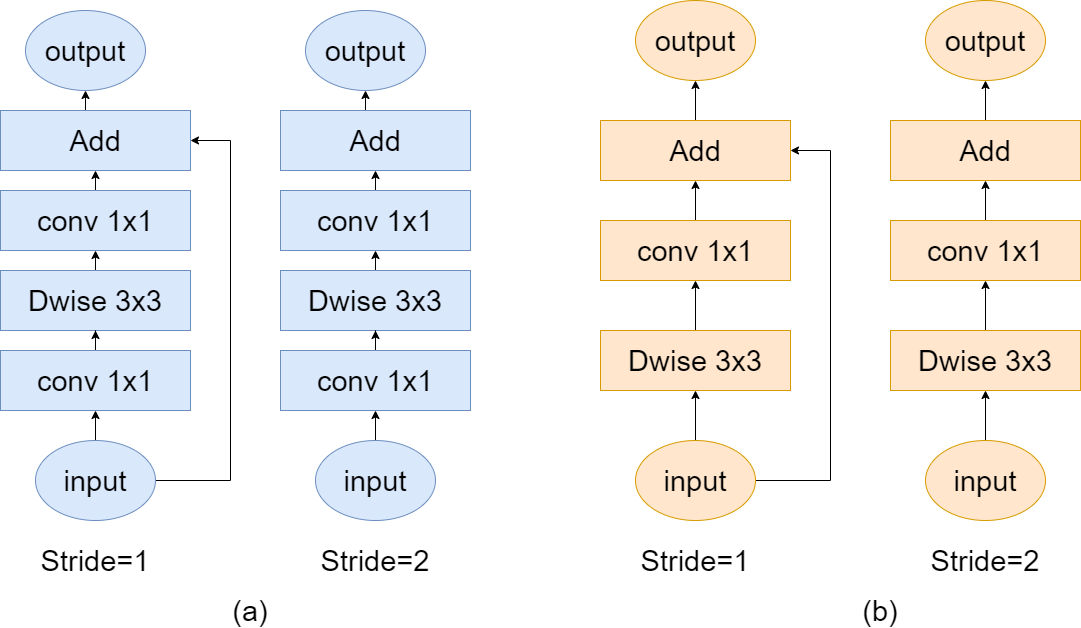}}
	\caption{(a) The basic MobileNetv2 block, and (b) the proposed one}
	\label{fig:newmobilenetv2}
\end{figure}

To make YOLO-v2 fusion ready, we replace the basic convolutional layer with the MobileNetv2\cite{mobilenet} block that combines one depthwise, two pointwise convolutions, and skip connection layers as shown in Fig.~\ref{fig:newmobilenetv2} (a). In this work, inspired by \cite{radosavovic2020designing} that the expansion factor used in MobileNetv2 is not a must, we remove the first pointwise layer of MobileNetv2 as shown in Fig.~\ref{fig:newmobilenetv2} (b). Other model compression approaches can also be applied to reduce model size. This step can be skipped if the input model is near fusion-ready. Our RC-YOLOv2 is constructed by stacking the blocks shown in Fig.~\ref{fig:newmobilenetv2} (b).

\begin{figure}[t]
	\centering{\includegraphics[width=0.45\textwidth]{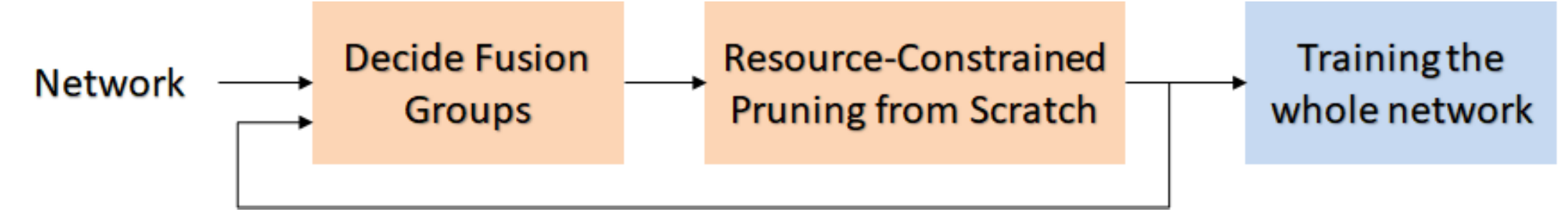}}
	\caption{The buffer size constrained structure morphing}
	\label{fig:moprhnet}
\end{figure}
\begin{figure}[t]
	\centering{\includegraphics[width=0.45\textwidth]{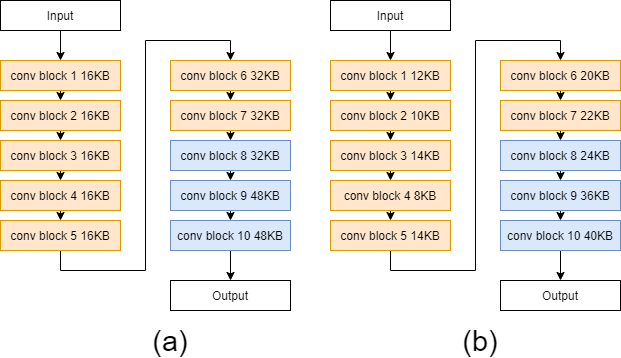}}
	\caption{An RCNet example at the first iteration,: (a) before RCNet, and (b) after RCNet. Best view in colors.}
	\label{fig:rcnetex1}
\end{figure}\begin{figure}[t]
	\centering{\includegraphics[width=0.45\textwidth]{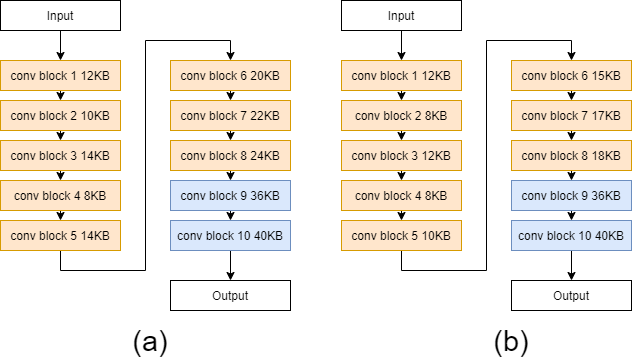}}
	\caption{An RCNet example at the second iteration: (a) before RCNet, and (b) after RCNet. Best view in colors.}
	\label{fig:rcnetex2}
\end{figure}
\begin{figure*}[t]
     \centering
     \subfigure[]{
         \centering
         \includegraphics[width=0.65\textwidth]{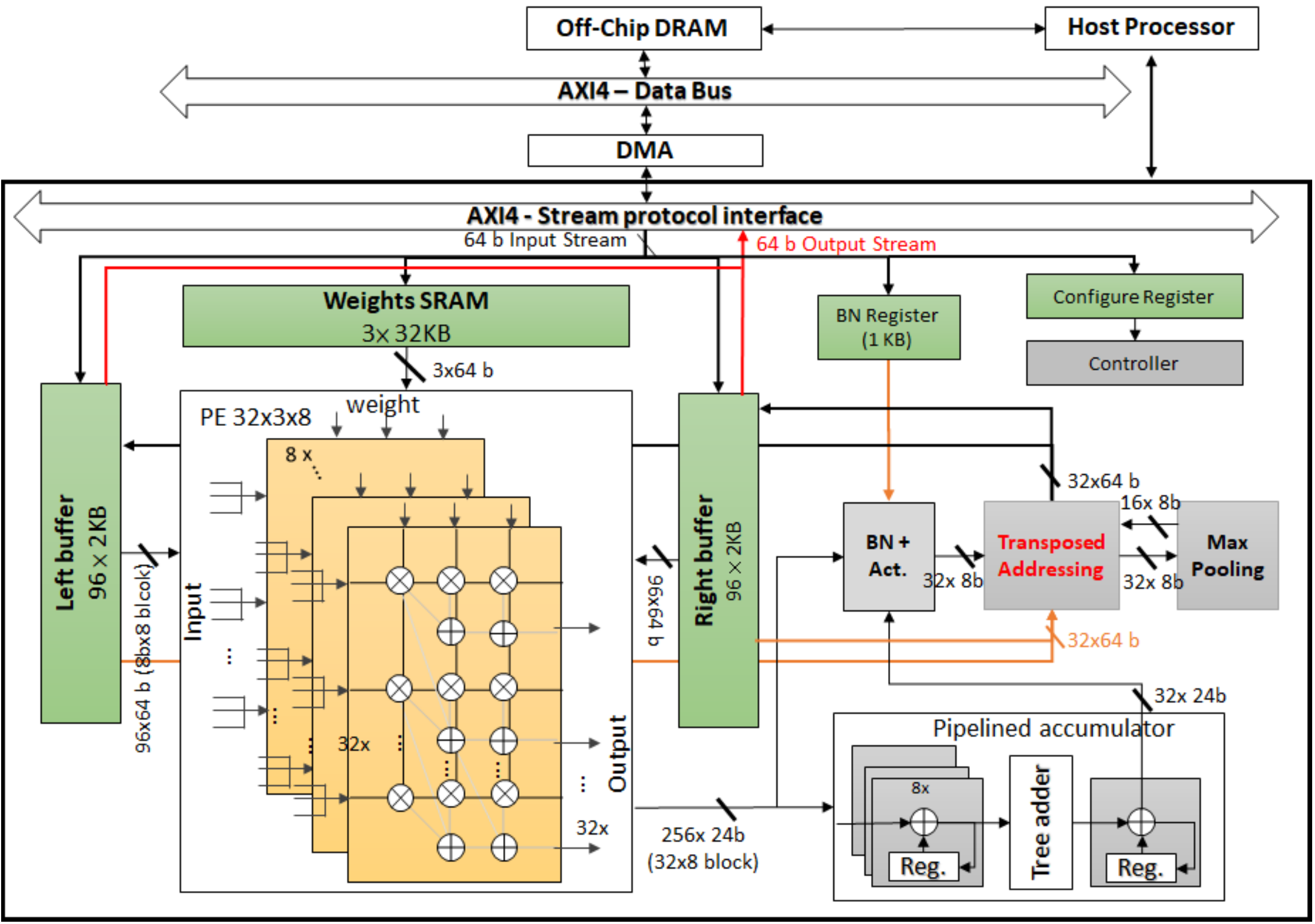}}
         \subfigure[]{
         \centering
         \includegraphics[width=0.3\textwidth]{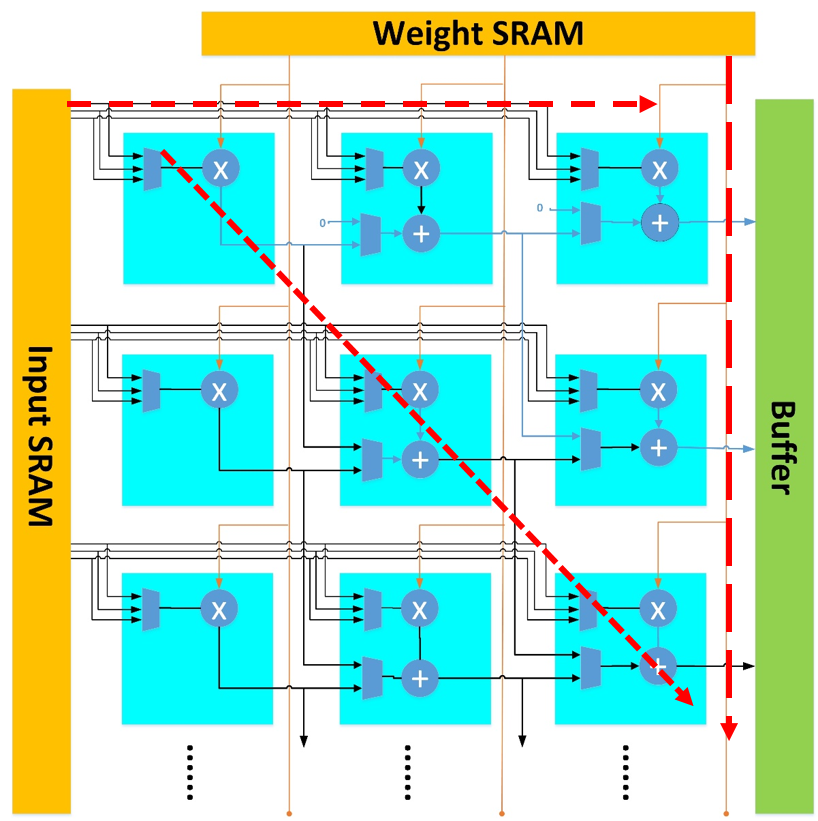}}
    \caption{(a) The proposed system architecture, and (b) its dataflow}   
	\label{3-top}
\end{figure*}


 \begin{figure*}[h]
	\centering{\includegraphics[height=100mm]{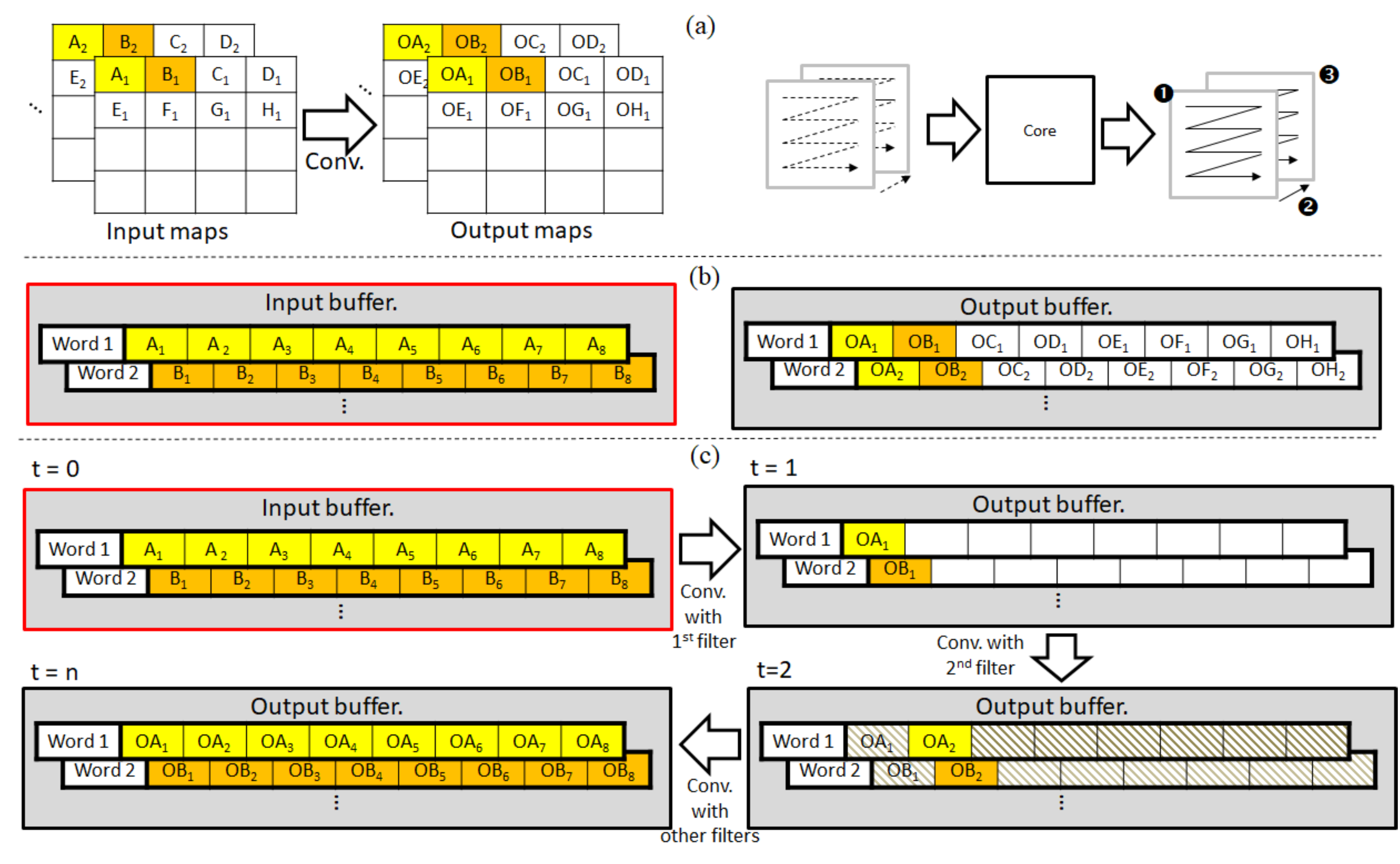}}
	\caption{(a) Illustrations of input and output maps and output generating order of output data reuse (b) The data order for the input and output buffers without any transposed addressing. (c) The data order for the input and output buffers with transposed addressing by the write-masking in SRAM.}
	\label{4-2-wS2CA}
\end{figure*}
\begin{figure}[t]
	\centering{\includegraphics[height=45mm]{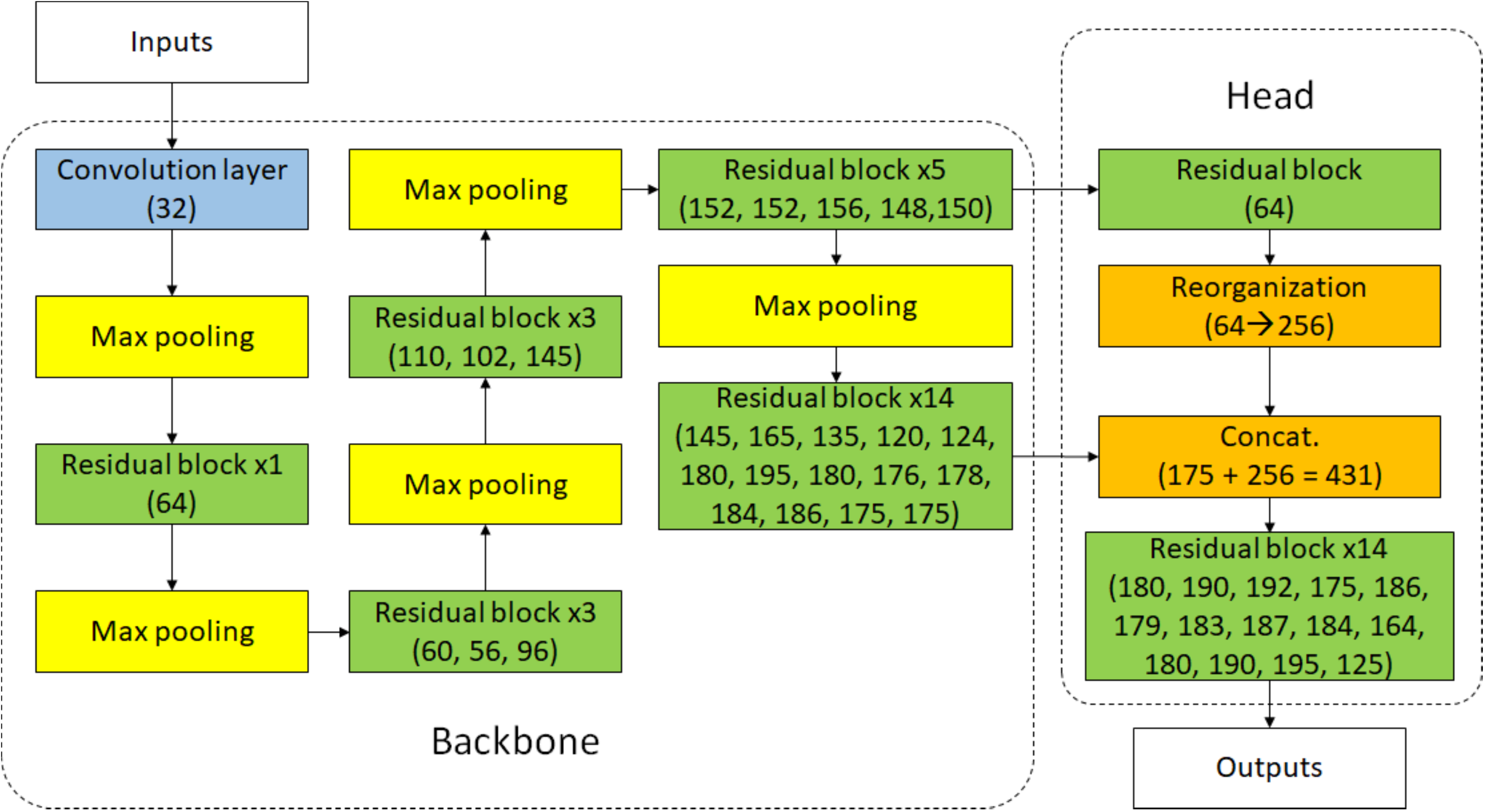}}
	\caption{Model structure in the morphed YOLO-v2. Numbers in each block represent the output channel. Structures in residual blocks are shown as Fig.~\ref{fig:residual}.}
	\label{fig:model}
\end{figure}

\begin{figure}[t]
	\centering{\includegraphics[height=60mm]{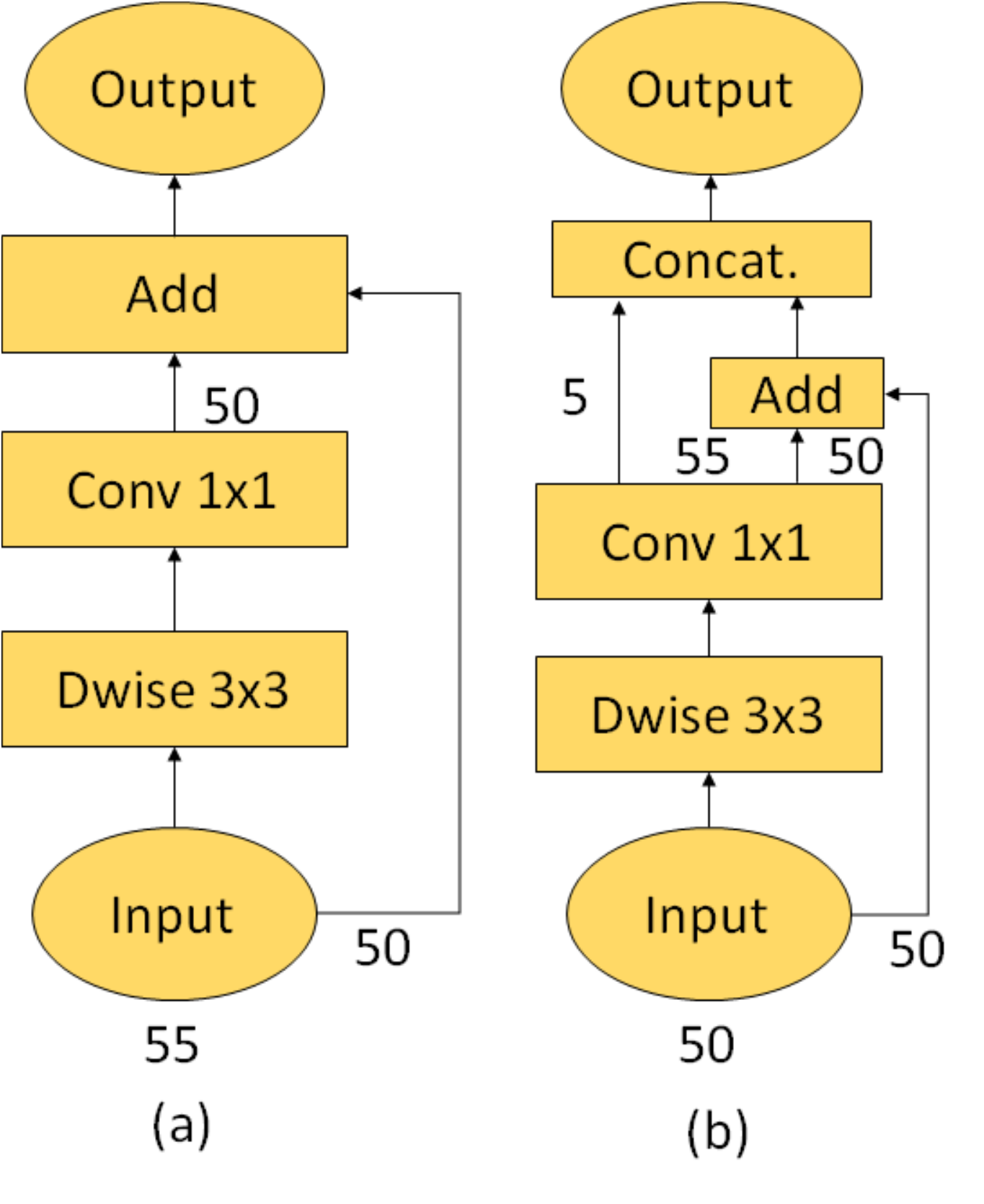}}
	\caption{Structures in residual blocks. (a) The channel number of layer $L-1$ is larger than layer $L$. (b) The channel number of layer $L-1$ is smaller than layer $L$.}
	\label{fig:residual}
\end{figure}

\subsection{Buffer size constrained structure morphing}
Even after lightweight conversion, directly fusing multiple layers as in \cite{blockconv} cannot maximize the number of fused layers due to the uneven distribution of per layer size. To solve this, we propose a new pruning procedure inspired by \cite{gordon2018morphnet} to maximize the number of fused layers while remaining within the weight buffer size set by designers. Fig.~\ref{fig:moprhnet} shows the detailed procedure. We first decide the group partition for fusion by analyzing the model from input to output. If the size of a layer exceeds the size of the available weight buffer, the fused layer group ends from its previous layer and we start a new fusion group from this layer. Our goal is to create a network in which the total weight size of each fusion group is close to the weight buffer size $B$, and to fuse more layers to avoid external memory traffic for intermediate data.

\subsubsection{Formulation}

\noindent The goal of training any network is to minimize a loss$\mathrm{\ }L$:
\[{\mathop{\mathrm{\ min}}_{\theta } L\left(\theta \right)\ }\] 
where $\theta $ denotes all trainable parameters of the network and\textit{ }$L$ is a loss measuring how well a network is and depends on the target task. Let $O$ denotes the network after model conversion. In the first step, we partition the network into several fusion groups$\mathrm{\ }G=\{G_1,G_2,\bullet \bullet \bullet ,G_k\}$. As mentioned above, the fusion strategy is from input to output. In this step, we allow the total weight size of each fusion group to exceed the weight buffer size in a certain range (50\% for example in this work). In the second step, the total weight size of the fusion group will be slimmed to be smaller than the weight buffer size. At the same time, the new network $O'$ learns the optimal structure with these fixed fusion groups. The formulation is as below 

\begin{equation}
   \begin{aligned}
      \mathrm{\ \ }O^{'} ={\mathop{argmin}_{F\left(G_i\right)\le B,\ \ \ \forall i=1\sim k} {\mathop{\mathrm{\ min}}_{\theta } L\left(\theta \right)\ }\ }\ \   
  \end{aligned}
   \label{eq:f1}
\end{equation}

\noindent where $F$\textit{ }represents the total weight size of each fusion group. To reduce the weight size of each fusion group, we use L1 regularization to train the scaling factor $\gamma$ in BN that is applied after every convolutional layer and set the model size as a constraint. Therefore, the optimization problem is as below:
\begin{equation}
                      {\mathop{\mathrm{\ min}}_{\theta } L\left(\theta \right)\ }+\lambda \delta (\theta )                   
   \label{eq:f2}
\end{equation}

\noindent where $\lambda$ denotes the regularization parameter, and $\delta \left(\theta \right)$ denotes the regularization term constrained by the weight size. Assume the weight size of a convolutional layer is$\mathrm{\ }S_l$. We can rewrite the weight size as:
\begin{equation}
F\left(\mathrm{layer\ L}\right)\mathrm{=}S_l\sum^{C^L_{in}-1}_{i=0}{A_{L,i}}\sum^{C^L_{out}-1\ }_{j=0}{B_{L,j}} 
\end{equation}
where $A_{L,i}$($B_{L,j})$ is the indicator function that is zero if the \textit{i}-th input (\textit{j}-th output) of the layer $L$ is zeroed out. ${C^L_{in}}$ and ${C^L_{out}}$ are the input and output channel for the $L-th$ layer, respectively. Therefore, the regularization term of the layer $L$ becomes
\begin{equation}
\begin{split}
\delta \left(\theta ,layer\ L\right)=\ S\sum^{C^L_{in}-1}_{i=0}{\left|{\gamma }_{L-1},i\right|}\sum^{C^L_{out}-1\ }_{j=0}{B_{L,j}}+ \\ S\sum^{C^L_{in}-1}_{i=0}{A_{L,i}}\sum^{C^L_{out}-1\ }_{j=0}{\left|{\gamma }_L,j\right|} 
\end{split}
\end{equation}
\\
Then, the whole regularization term becomes
\begin{equation}
\delta \left(\theta \right)=\sum^N_{L=1}{\left(\theta ,layer\ L\right)} 
\end{equation}
where N represents the number of layers. After training, we prune the channel with the smallest $\gamma$ value in each fusion group to satisfy the weight buffer size constraint. 

To avoid a long training time, we adopt the pruning-from-scratch method from \cite{wang2020pruning}. Instead of $\gamma$ values, all other trainable parameters like weights in the convolutional layer are frozen for training and set as random values. We only train the $\gamma$ value under fixed random weights. Therefore, (\ref{eq:f1}) becomes:
\begin{equation}
  \begin{aligned}
O^{'}={\mathop{argmin}_{F\left(G_i\right)\le B,\ \ \ \forall i=1\sim k} {\mathop{\mathrm{\ min}}_{\gamma } L\left(\gamma \right)\ }\ }\ \ 
\end{aligned}
\end{equation}
\noindent and (\ref{eq:f2}) becomes:
\begin{equation}
{\mathop{\mathrm{\ min}}_{\gamma } L\left(\gamma \right)\ }+\lambda \delta (\gamma )
\end{equation}
\noindent These can speed up the training procedure significantly without performance degradation. We only have to train the entire network with all trainable parameters once when we finally obtain the low memory traffic model.

 This procedure is done iteratively. Once going through one iteration, the total model size of the network will become smaller. The procedure can be stopped at any time when meeting the requirements or when encountering a dramatic accuracy drop. In which, to avoid the new network bounded by its original shape, at the first few iterations, we use uniform channel width scaling to scale the entire network to its original model size after the second step.

\subsubsection{Algorithm}
\begin{algorithm}
\caption{RCNet: Resource-constrained network fusion and pruning}
\label{alg:cap}
\KwIn{Initial network, weight buffer size $B$, extra buffer size range $m$}
\KwResult{Result network $O^{'}$}    
Do model conversion and get a fusion ready network $O$\;
Decide the group partition $G$ from input to output for layer fusion. Total weight size of the group $F(G)\leq (1+ m)\times B$\;
Train the network to find $O^{'}={\mathop{\mathrm{\ argmin}}_{\gamma } L\left(\gamma \right)\ }+\lambda \delta (\gamma )$\ for suitable $\gamma$\;
Find the new network $O^{'}$ induced by $\gamma^\ast$ such that $F(G) \leq B$ \;
At the first few iterations of the training, scale the network back to its original size\;
Repeat from Step 2 for many times as desired\;
\end{algorithm}

The whole algorithm of RCNet is shown in Algorithm \ref{alg:cap}. The first step "model conversion" is suggested to if the model size is too large for fusion. The conversion method in Section II-B can be replaced with other model compression methods as well. Fig. \ref{fig:rcnetex1} and \ref{fig:rcnetex2} shows an example of how the RCNet works with the buffer size in each layer set for illustration purposes. We assume the weight buffer size to be 100 KB, and the extra buffer size range $m=50\%$ in this case. In Fig. \ref{fig:rcnetex1} (a), the initial fusion group after step 2 will have two groups with sizes of 144 KB and 128 KB, respectively. After applying RCNet, the total weights of the two fused layer groups will be pruned to 100 KB as in Fig. \ref{fig:rcnetex1} (b). Note that the illustration does not scale the channels back so that the model size of the overall network will be smaller after each iteration. Then in the second iteration as in Fig. \ref{fig:rcnetex2} (a), the first fusion group will expand one more layer after step 2, and the weight size of the two groups will be 124 KB and 76 KB. After applying RCNet, the weight size of the first group is pruned to be 100 KB as in Fig. \ref{fig:rcnetex1} (b). With the above procedure, we can fuse as many layers as possible.  In our experiment, we set $m=50\%$, and just need to run the iteration one or two times to get the desired result.

\subsubsection{Hardware oriented fusion guidelines}
In the RCNet, the layers to be fused depend on the model structure and weight buffer size. In general, pooling or stride layers will affect hardware utilization significantly due to much smaller input. However, the first input layer will not be affected by this significantly due to its larger input. Besides, the group partition within a residual block will cause extra data access due to the shortcut connection. Thus, based on the above observations, we combine the following hardware-oriented fusion guidelines with the RCNet to make the hardware fully utilized and reduce unnecessary traffic for YOLOv2.
\begin{itemize}
\item \textit{The first layer with downsampling shall be fused with other layers to increase utilization.}
For computer vision tasks, the image input comprises only three input channels. The first convolutional layer usually follows by a pooling layer or uses strides to reduce the computation cost. To improve the benefits of layer fusion in the first fusion group, we will ignore the downsampling of the first layer and fuse the first layer with others according to buffer size constraints. Moreover, due to only three channels in the first layer, the tile size can be maximized and stored in a unified buffer. Thus, PE utilization can still be kept high even after the pooling layer.
\item \textit{A group of fused layers shall have no more than two downsampling layers.}
The downsampling layers such as the pooling layer or stride will reduce the feature map size. However, too many pooling layers will cause too small feature maps to fully utilize parallel PEs. Thus, as a design tradeoff for our current design, we limit no more than two downsampling layers. This also applies to the convolution with strides.
\item \textit{A residual block shall be in the same group of fused layers.}
The input data of the first layer in a residual block needs to be stored due to the skip connection. To avoid DLAs accessing the skip input from DRAM repeatedly, we prefer that all layers in a residual block must be in the same group of fused layers.

\end{itemize}

Of the three guidelines, the first two are for hardware utilization. If they are not met, the hardware utilization will become lower and thus result in lower throughput. The final one is for memory bandwidth. If we cannot fuse the layers within the residual connections, extra external DRAM access will be needed. The impact depends on the size of the feature input in that residual block. In general, the data access of the feature input will be one time than that with fusion.

\section{System Architecture}
\subsection{Overview}

Fig.~\ref{3-top} shows the architecture of the proposed DLA based on our prior design\cite{VWA}. This design is a typical systolic array type with tile-based scheduling. It has a weight buffer for storing fusion weights and a unified buffer with left and right buffers for storing the feature maps. One PE block has an n×3 MAC array design. The PE block has n inputs broadcasted horizontally in the same input column vector, and three weights broadcasted vertically in the same weight column vector to optimize for 3×3 convolutions. The multiplication results are summed along the diagonal direction. Finally, the partial sum of PE blocks will be accumulated at the accumulator to generate the output. To support the HD size object detection, this DLA core has 768 multiplier accumulator (MAC) units, which are split into 8 PE blocks. Each PE block consists of a 32x3 MAC array with 32 feature inputs and three weight inputs.

\subsection{Hardware support for layer fusion group}
To support layer fusion, the original design has two major architecture changes: a larger weight buffer for fusion group weights and a unified buffer for fusion group execution. Besides, the nonoverlapped tile processing from\cite{blockconv} is also used in the computing flow. The weight buffer size is set to 96 KB after several experiments as shown in the result section. Nonoverlapped tiling partitions an input into nonoverlapped tiles and processes them with constant boundary extension or zero padding. This can help to avoid data dependency between tiles and make layer fusion easier. 

However, for a small tile size, the accuracy loss is greater. In this work, the tile size is determined by the on-chip feature buffer size, as opposed to the adhoc fixed value in \cite{blockconv}. For a fusion group, we first assume the required input feature map size ($map$) and then calculate the output feature map size for all layers in a fusion group based on the channel number and buffer size, i.e. $map/Pooling~Factor\times channels = Buffer~Size$. Then we select the smallest map size value as the input map size of this group to ensure that such size is smaller than the buffer size for each layer. Assume $Map~Size = Tile~Width \times Tile~Height$. Then, the width of the tile is set to be the same as the width of the feature map to avoid padding on the left and right boundaries of a tile. The top and bottom boundaries of a tile use boundary extensions for the nonoverlapped tile processing. Then the tile height can then be set as the maximum allowable value. This tile processing is applied to the complete model.

The layer fusion execution in nonoverlapped tile processing is simply layer-by-layer processing within a fusion group, but all intermediate data is from the internal buffer rather than DRAM. Thus, the input and output buffers in the original design are now acted as a ping-pong buffer, which is denoted as the unified buffer as the left and right buffers shown in Fig.~\ref{3-top}. When the left buffer acts as the input buffer, the right buffer will act as the output buffer. Then, after the convolution has been completed, their roles as input and output are switched. Thus, all intermediate data are from the local buffers.

However, the input buffer is addressed along the spatial dimension, whereas the output buffer is addressed along the channel dimension. Because of this addressing inconsistency, it is inconvenient to use a direct I/O merging buffer, which has not been addressed in  \cite{blockconv}. To address this issue, one convenient solution is to use the write-masking capability of the on-chip SRAM to reorder the output data to match the desired input data.

This transposed addressing is illustrated in Fig.~\ref{4-2-wS2CA}. To support such access without extra overhead, we divide the data words into eight banks (e.g. A1~A8 in one bank, B1~B8 in another bank in Fig.~\ref{4-2-wS2CA} (b)) and use the byte write capability of the SRAM to write OA1~OH1 to different banks as in Fig.~\ref{4-2-wS2CA} (c). This can reorder the output to be the expected input order. This helps smooth layer fusion to avoid complex control and provide enough data amount.

\section{Experimental Result}
\subsection{Resource constrained YOLOv2 (RC-YOLOv2) for HD size object detection}
The RCNet is applied to our target HD size object detection, YOLO-v2 \cite{YOLOv2}. This paper adopts the original YOLO-v2 model as our baseline, which is trained with the Pascal VOC 2007 + 2012 dataset with 74.23\% mAP evaluated on the Pascal VOC 2007 dataset\cite{pascal}. The training uses SGD with weight decays. For the learning rate, we use the warm-up strategy from zero to 0.1 in the early epochs and then the step decay policy until 0.0001. We use L1 loss to regularize the scaling factor $\gamma$ in BN. The final model is pretrained on ImageNet and then trained on this dataset. 

The baseline model has 55.6M model parameters and needs 98 MB feature map I/O per inference, which will be 2.9 GB/s for 30 frames per second (FPS). With the proposed approach, the new model, RC-YOLOv2, is shown as Fig.~\ref{fig:model}. We can reduce the model parameter to 1.014 M with 72.12\% mAP and 5.01 MB feature map I/O under the constraint of 96 KB weight buffer size. The model size is reduced by 98.7\% (55.6M to 1.014M) and 73.3\% (3.806M to 1.014M) when compared to the baseline model before and after model conversion, respectively.   

The channel number of the shortcut and 1x1 convolution path will be inconsistent for the summation of the residual block after the RCNet pruning. To solve this issue, our higher priority is to keep the channel number from the 1x1 convolution and sum with the block input of the same number of channels. We discard extra channels from the block input as shown in Fig. ~\ref{fig:residual} (a). If the channel number of the block input is smaller than the convolutional output, extra channels from the convolutional output are output directly as in  Fig. ~\ref{fig:residual} (b).

The RC-YOLOv2 is also trained and tested on an HD size dataset, IVS\_3cls \cite{Tsai} for object detection on road traffic, as shown in Table \ref{tab:IVS}. The input size is 1920x960. The original model has 88.2\% mAP, 55.6M model size, and 131.62 MB feature map I/O. With the proposed model conversion, the model size is significantly reduced, but the external I/O still needs 130.65MBs. A naive fusion for 100 KB weight buffer size only fuses a small fraction of layers and still needs 80.45 MB I/O. With RCNet, the result model can achieve 80.81\% mAP, 1.76M model parameters, and 21.15MBs feature map I/O. Further quantization to 8-bit does not affect accuracy. The accuracy drop can be recovered by pretraining on ImageNet before training on this dataset. The proposed RCNet can be applied to other tasks as well, like semantic segmentation as shown in Table~\ref{tab:deeplab} and image classification as shown in Table \ref{tab:VGG}.



\begin{table}
\centering
\caption{Ablation study of RC-YOLOv2 on IVS\_3cls for 100KB weight buffer size}
\label{tab:IVS}
\begin{tabular}{l|l|llll}
                 & YOLOv2 &        &       &       &        \\ 
\hline
Conversion Only? &        & {\checkmark}      & {\checkmark}    &       &        \\
Naive Fusion?    &        &        & {\checkmark}     &       &        \\
RCNet?           &        &        &       & {\checkmark}     & {\checkmark}     \\
Quantization?    &        &        &       &       &{\checkmark}     \\ 
\hline
mAP              & 88.2   & 84.3   & 84.3  & 80.81 & 80.02  \\
FLOPs (G)        & 625    & 80.2   & 80.2  & 38.69 &        \\
Model size (M)   & 55.66  & 3.8    & 3.8   & 1.76  &        \\
Feature I/O(MB)          & 131.62 & 130.65 & 80.45 & 21.55 &       
\end{tabular}
\end{table}

\begin{table}
\centering
\caption{Ablation study of DeepLabv3\cite{chen2017rethinking} on the PASCAL VOC 2012 dataset\cite{everingham2015pascal} for 100KB weight buffer size}
\label{tab:deeplab}
\begin{tabular}{l|l|llll}
                 & DeepLabv3 &        &       &       &        \\ 
\hline
Conversion Only? &        & {\checkmark}      & {\checkmark}    &       &        \\
Naive Fusion?    &        &        & {\checkmark}     &       &        \\
RCNet?           &        &        &       & {\checkmark}     & {\checkmark}     \\
Quantization?    &        &        &       &       &{\checkmark}     \\ 
\hline
mIOU                & 70.5      & 68.8  & 68.8  & 67.1 & 65.9 \\
FLOPS   (G)         & 51.29     & 23.28 & 23.28 & 4.86 &      \\
Model   size (M)    & 39.64     & 9.11  & 9.11  & 2.2  &      \\
Feature I/O (Mbyte) & 52        & 50.2  & 27.31 & 6.36 &          
\end{tabular}
\end{table}

\begin{table}
\centering
\caption{Ablation study of VGG16\cite{simonyan2014very} on the ImageNet dataset\cite{russakovsky2015imagenet} for 200KB weight buffer size}
\label{tab:VGG}
\begin{tabular}{l|l|llll}
                 & VGG16 &        &       &       &        \\ 
\hline
Conversion Only? &        & {\checkmark}      & {\checkmark}    &       &        \\
Naive Fusion?    &        &        & {\checkmark}     &       &        \\
RCNet?           &        &        &       & {\checkmark}     & {\checkmark}     \\
Quantization?    &        &        &       &       &{\checkmark}     \\ 
\hline
Top5        & 92.5 & 90.2 & 90.2 & 89.7 & 89.5 \\
FLOPS   (G)      & 30.74     & 5.42      & 5.42      & 3.89      &           \\
Model   size (M) & 15.23     & 4.45      & 4.45      & 2.53      &           \\
Feature I/O (Mbyte)      & 48.6      & 48.25     & 16.32     & 7.68      &          
\end{tabular}
\end{table}

\begin{figure}[bt]
	\centering{\includegraphics[width=0.45\textwidth]{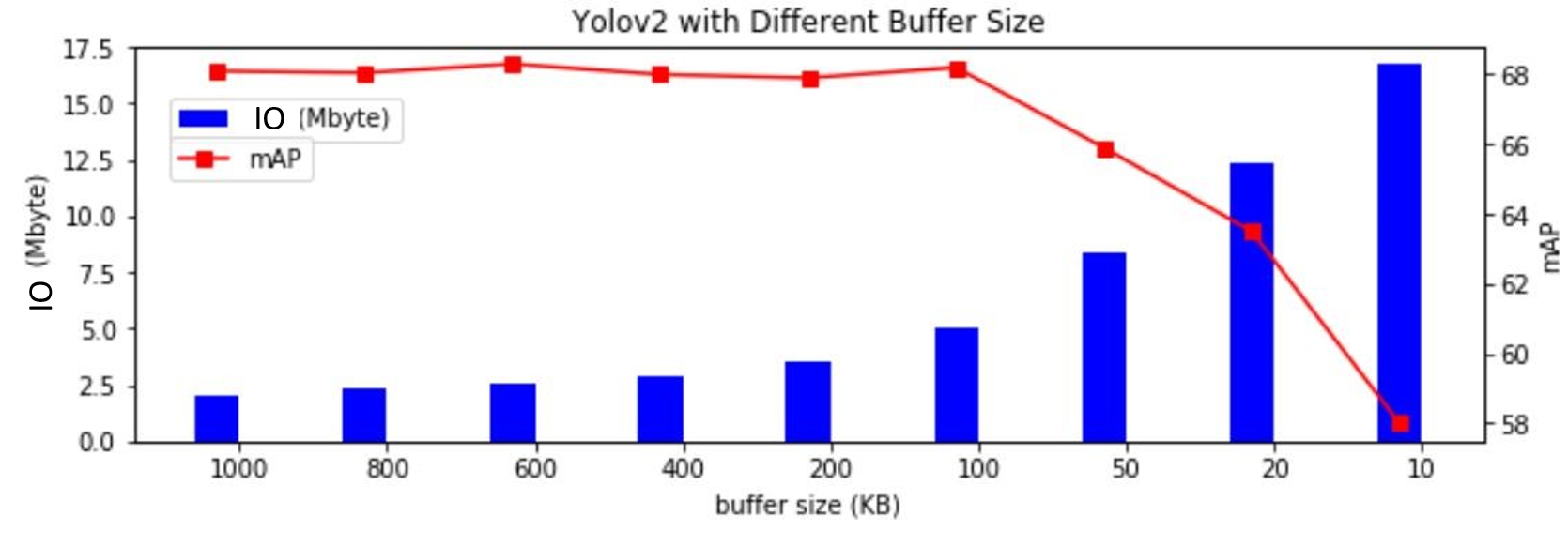}}
	\caption{RC-YOLOv2 under different weight buffer sizes.}
	\label{fig:rcbuffer}
\end{figure}

\begin{figure}[bt]
	\centering{\includegraphics[width=0.45\textwidth]{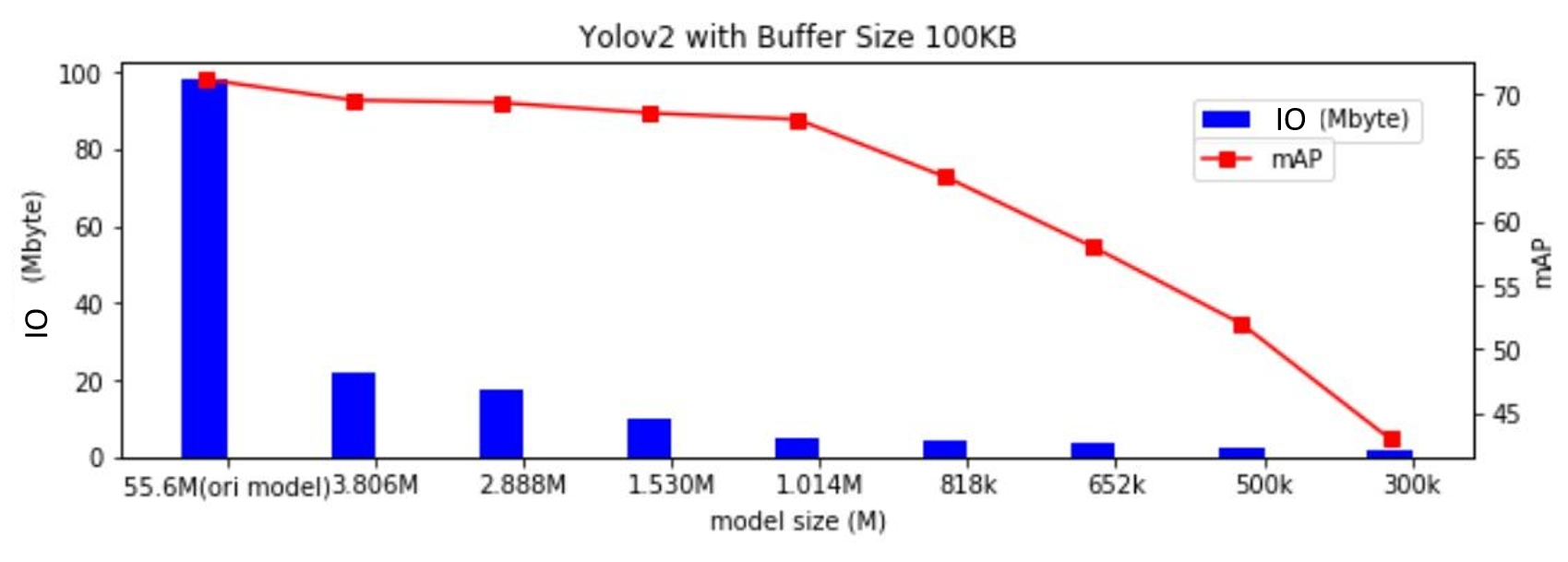}}
	\caption{RC-YOLOv2 for different final model sizes under 100 KB weight buffer size.}
	\label{fig:rcbufferm}
\end{figure}

\begin{figure}[t]
	\centering{\includegraphics[height=40mm]{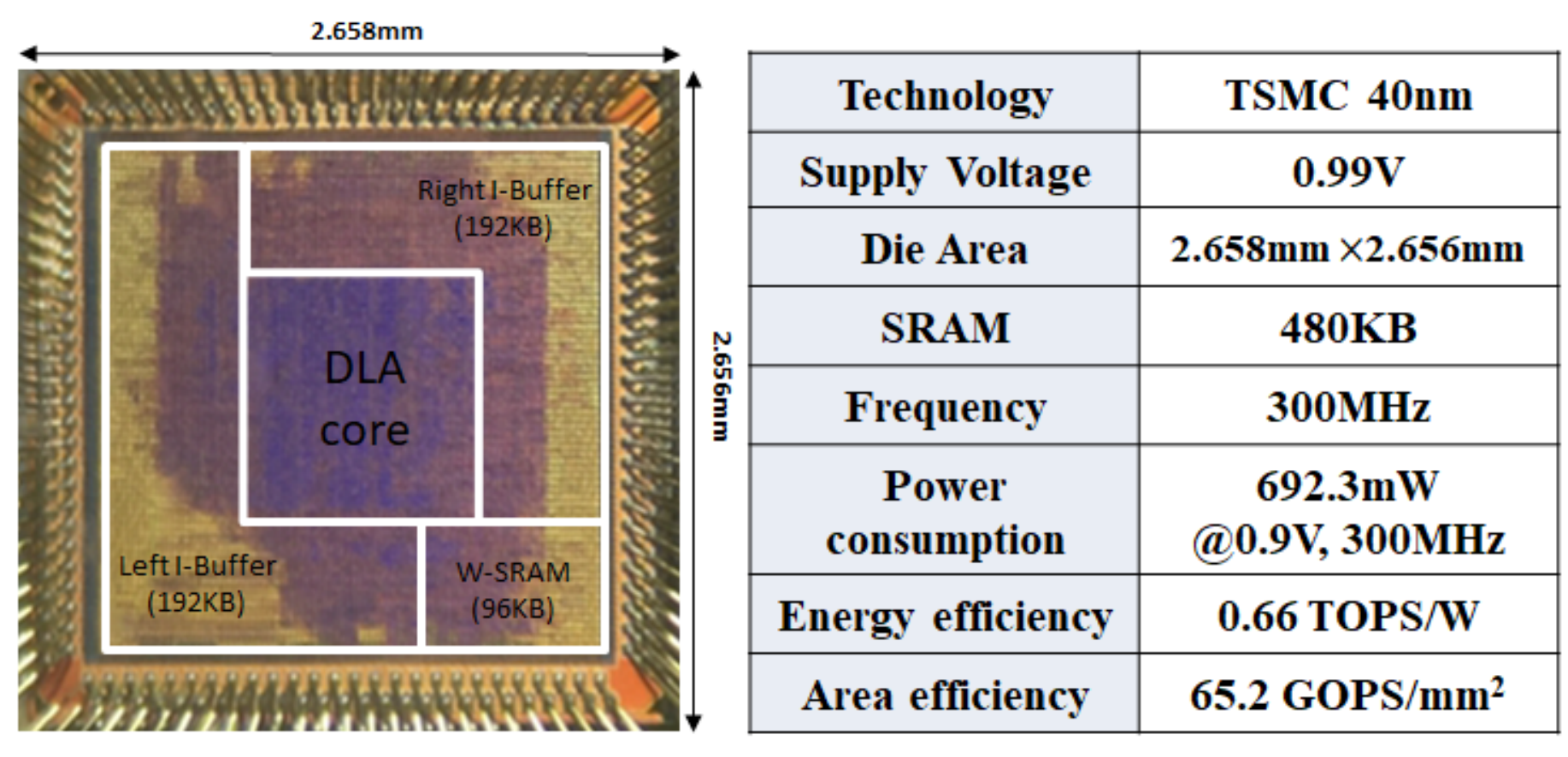}}
	\caption{Die photo and implementation results.}
	\label{8-implement}
\end{figure}
 \begin{figure*}[htb]
	\centering{\includegraphics[height=50mm]{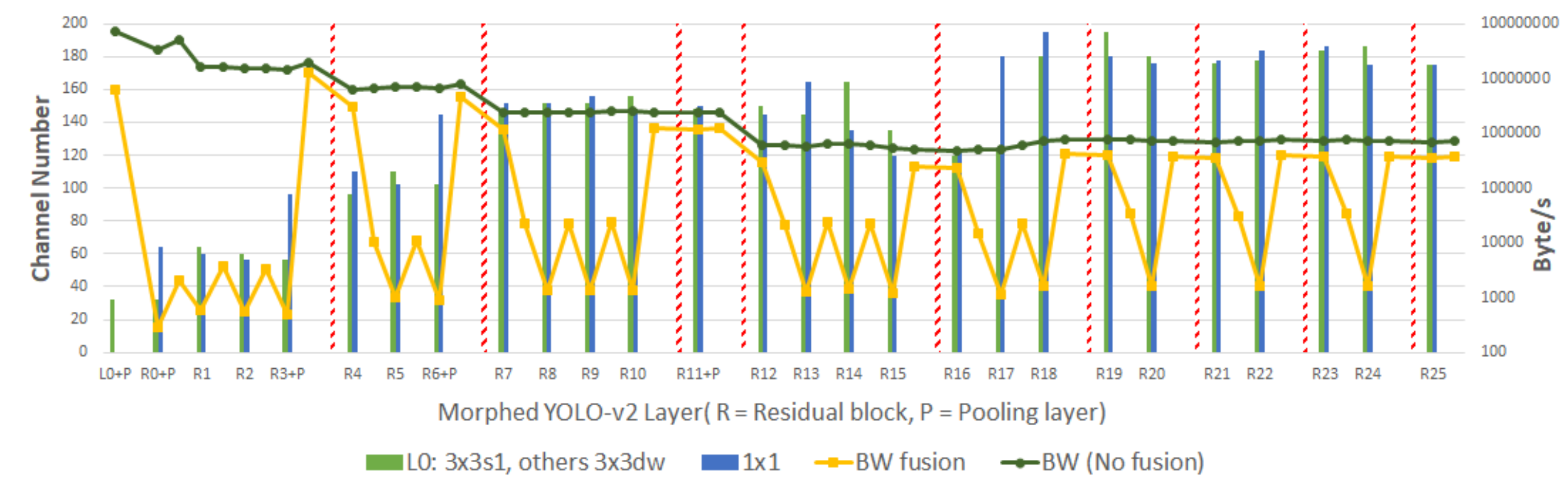}}
	\caption{The external data amount of each layer for the backbone of the RC-YOLOv2.}
	\label{9-modelresult}
\end{figure*}

In RCNet, an important constraint is the size of the weight buffer. Different buffer sizes will lead to different network structures. Overall speaking, a larger buffer will need smaller feature I/O and has higher accuracy since the constraint caused by the buffer size is smaller. However, a larger buffer size means a higher cost for DNN accelerator design. To study this effect, 
Fig. \ref{fig:rcbuffer} shows the weight buffer size effect on RC-YOLOv2 for the target total model size around 1M. Feature I/O goes higher with a smaller buffer size. When the buffer size is under 100 KB, the mAP drop will be significant. Therefore, in this work, we select the weight buffer size around 100 KB, which is 96 KB based on the selected PE numbers. Fig. \ref{fig:rcbufferm} shows the results for different final model sizes under 100 KB weight buffer. The size of the network can be reduced to about 1M within 3\% mAP drop. We can also obtain similar results for different weight buffer sizes. Therefore, we select 1M as our final model size target.

\subsection{Hardware Implementation Result}
Fig.~\ref{8-implement} shows the chip photograph and performance summary, which is fabricated with a TSMC 40 nm CMOS process. It occupies 4.56 mm$^2$ with 480 KB
SRAM, including 384 KB unified buffer and 96 KB weight buffer. The peak performance is 460.8 GOPS at 300MHz for full PE utilization. This chip can execute this object detection model at 30FPS for 1280$\times$720 HD images and 20 FPS for 1920$\times$1080 full HD images. The core power consumption is 692.3mW, measured by running the RC-YOLOv2.





Fig.~\ref{9-modelresult} shows the channel number and data bandwidth of each layer for the RC-YOLOv2 model targeted to HD (1280$\times$720) image input. This figure also shows the groups of fused layers separated by red dashed lines, which are usually at the pooling layer. Group 1 comprises a 3x3 convolution with pooling and two residual blocks with pooling. This group follows the above guideline on the first layer since the image input only has three input channels that will cause low hardware utilization. Besides, the weight size of the first group is small, and the map size is large. Thus, more fused layers in this group can reduce a large amount of bandwidth in this model. In addition, the second and the fourth fusion groups are separated by only one pooling layer since more pooling layers in these groups will cause low hardware utilization. Other groups are decided by fusing as many layers as possible to fit the weight buffer size. The data bandwidth is shown as the yellow line chart in the figure, which includes all feature map I/O and weight access. With the group fused layers, the layer-by-layer external memory traffic can be reduced by 37\% - 99\% for different layers.



\begin{table}[b]
\centering
\caption{Memory traffic and energy comparison for 30FPS. Assume DDR3 DRAM energy consumption 70pJ/bit.}
\label{tab:membw}
\begin{tabular}{|l|l|l|l|l|l|} 
\hline
\multirow{2}{*}{Input size} & \multicolumn{2}{l|}{Bandwidth (MB/s)} & \multicolumn{2}{l|}{Energy(mJ)} & \multirow{2}{*}{Savings}  \\ 
\cline{2-5}
                            & Original & Proposed                   & Original & Proposed             &                           \\ 
\hline
416x416                     & 903      & 137                        & 506      & 77                   & 85\%                      \\ 
\hline
1280x720                    & 4656     & 585                        & 2607     & 328                  & 87\%                      \\
\hline
\end{tabular}
\end{table}

\begin{figure}[t]
	\centering{\includegraphics[height=60mm]{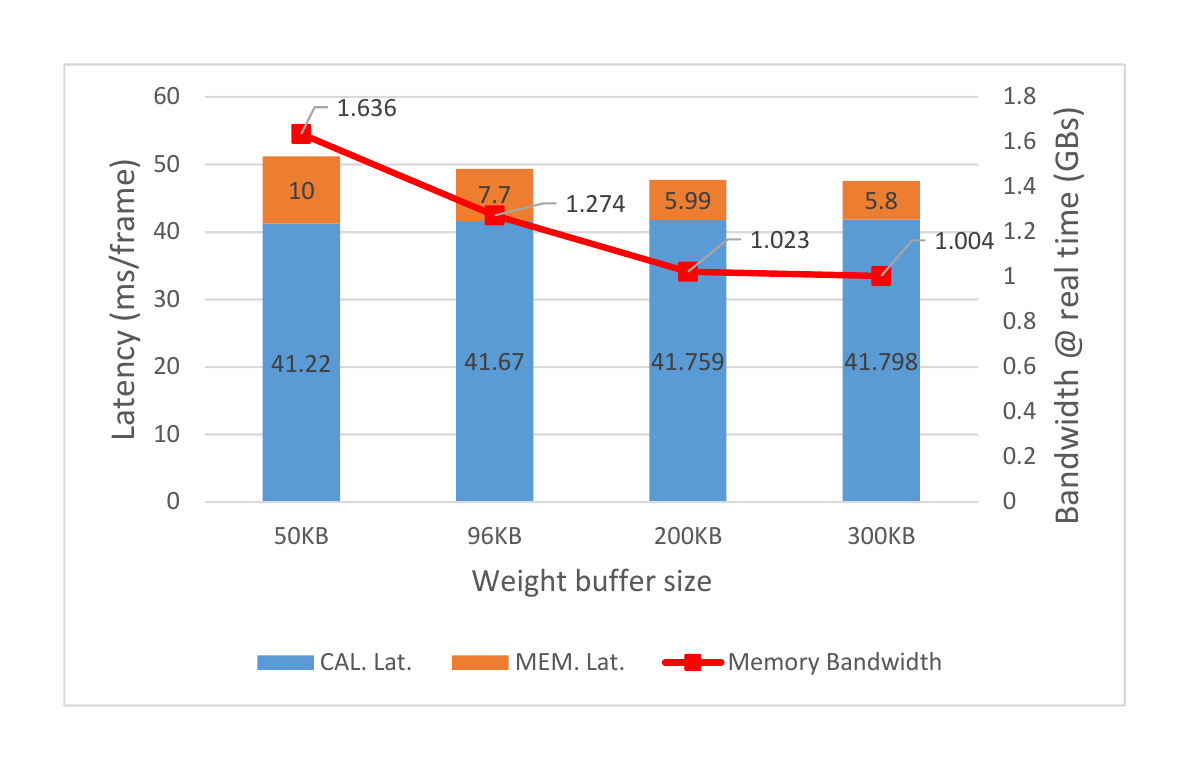}}
	\caption{The latency and memory bandwidth for different weight buffer sizes under design options with fusion and two 192KB unified buffers.}
	\label{17-weightbuffer}
\end{figure}

\begin{table*}[t]
	\centering
	\caption{Comparisons with other designs, where OD: object detection, GP: general purpose, and SR: super resolution.}
	\label{table:overview}
	\begin{tabular}{|l||c|c|c|c|c|c|c|}
		\hline
		           &Our Work& \cite{eyeriss} & \cite{eyerissv2} &  Envision\cite{envision} & \cite{MTK_fusion_isscc}  & SRNPU\cite{SRNPU_JETCAS} & THINKER\cite{thinker_2018jssc} \\\hline
		
		Technology & 40nm & 65nm& 65nm&28nm  & 7nm & 65nm & 65nm\\\hline
		Measurements & Chip &Chip&Post-layout&Chip&Chip&Chip & Chip\\\hline
		Layer fusion & Y  & - & -  & - & Y & Y & - \\\hline
		Task & OD  & GP & GP  & GP & GP & SR & GP \\\hline
		Supply Voltage (V) & 0.9 & 1.0 & 1.0 & 0.65 - 1.1   & 0.575 - 0.825  &1.1 & 1.2\\\hline
		Precision (bits) & 8,24 FXP &16 FXP& 8-20 FXP &4, 8, 16 Dyna.  & 8,16 FXP, 16 FP &  8,16 FXP & 8, 16 FXP \\\hline		
		PE number&768 & 168 &192 &-& -&- &1024\\\hline
		Clock rate (MHz)&300&200&200&200&290 - 880&200&200\\\hline
		\multirow{2}{3cm}{$^{a}$Peak Throughput (GOPS)}&460.8 &67.2 &153.6 &102 - 408&3604&232.1 &409.6\\
		&$^{c}$460.8&$^{c}$109.2&$^{c}$249.6&$^{c}$71.4 - 285.6&$^{c}$360.35&$^{c}$377.2&$^{c}$666.2\\\hline
		$^{b}$Area (KGE) (logic only)&1838&1176&2695&-&-&-&2950\\\hline
		Area (mm$^{2}$)&4.56&$^{e}$12.25&-&$^{e}$1.87&$^{e}$3.04&$^{e}$16&$^{e}$14.44\\\hline
		SRAM(KB)&480&181.5&192&144&2176&572&348\\\hline
		\multirow{2}{3cm}{$^{d}$Area eff. (GOPS/KGE)}&0.25&0.057&0.057&-&-&-&0.138\\
		&$^{c}$0.25&$^{c}$0.092&$^{c}$0.092&-&-&-&$^{c}$0.22\\\hline
		\multirow{2}{3cm}{$^{d}$Area eff. (GOPS/mm$^{2}$)}&101.05&5.485&-&54.5 - 218&1185& 14.5&28.36\\
		&101.05&$^{c}$8.914&-&$^{c}$38.1 - 152&\bf{$^{c}$207.35}&$^{c}$23.56&$^{c}$46.08\\\hline	
		$^{g}$Power (mW)&692.3&278&460.5&$^{i}$7.5 - $^{j}$300&174 - 1053&211&386\\\hline
		\multirow{2}{3cm}{Power eff. (TOPS/W)}&0.66&0.241&0.333&$^{j}$0.26 - $^{i}$10&3.42-6.83&1.1&1.06\\
		&\bf{$^{h}$0.66}&$^{h}$0.483&$^{h}$0.668&$^{hj}$0.27 - \bf{$^{i}$3.65}&$^{h}$0.48 - 0.502&$^{h}$2.6&$^{h}$3.06\\\hline
		\multicolumn{3}	{|l}{$^{a}$1 GMACS= 2 GOPS} & \multicolumn{5}	{l|}{$^{b}$The area is shown in terms of the size of kilo NAND2 gates (KGE).} \\
		\multicolumn{3}	{|l}{$^{c}$Technology scaling ($\dfrac{process}{40nm}$)} & \multicolumn{5}	{l|}{$^{d}$We take the theoretical performance to evaluate area efficiency fairly here.} \\
		\multicolumn{3}	{|l}{$^{e}$Core only size.} & \multicolumn{5}	{l|}{$^{f}$Chip size.} \\	
		\multicolumn{3}	{|l}{$^{g}$Core only power.} & \multicolumn{5}	{l|}{$^{h}$Normalized~power~efficiency $=$ power~efficiency $\times(\dfrac{process}{40nm})\times(\dfrac{Voltage}{0.9V})^2$.} \\
		\multicolumn{3}	{|l}{$^{i}$30\% - 60\% sparsity, 3 - 4bits, and 76 GOPS at 0.65V.} & \multicolumn{5}	{l|}{$^{j}$16 bits precision and 76 GOPS at 1.1V.} \\
		\hline		
	\end{tabular}
\end{table*}

Table \ref{tab:membw} shows data bandwidth comparison of the RC-YOLOv2. The proposed approach can save 85\%, and 87\% of memory traffic for image size 416 $\times$ 416 and HD (1280$\times$720) with 30 FPS, respectively, which is 6.5x and 7.9x reduction. Larger inputs will be benefited more from the fused layer. This required bandwidth is easily fallen within the range of DDR3 DRAM bandwidth (12.8 GB/s).

Fig.~\ref{17-weightbuffer} shows latency and memory bandwidth under different weight buffer size constraints to execute RC-YOLOv2 model on full HD images. With a larger buffer size, more layers can be fused for lower memory bandwidth, reducing 38\% bandwidth from 50 KB to 200 KB buffer. The reduction is saturated for 300 KB buffer size because it already has the maximum group of fused layers for our lightweight model.

\begin{figure}[t]
	\centering{\includegraphics[height=60mm]{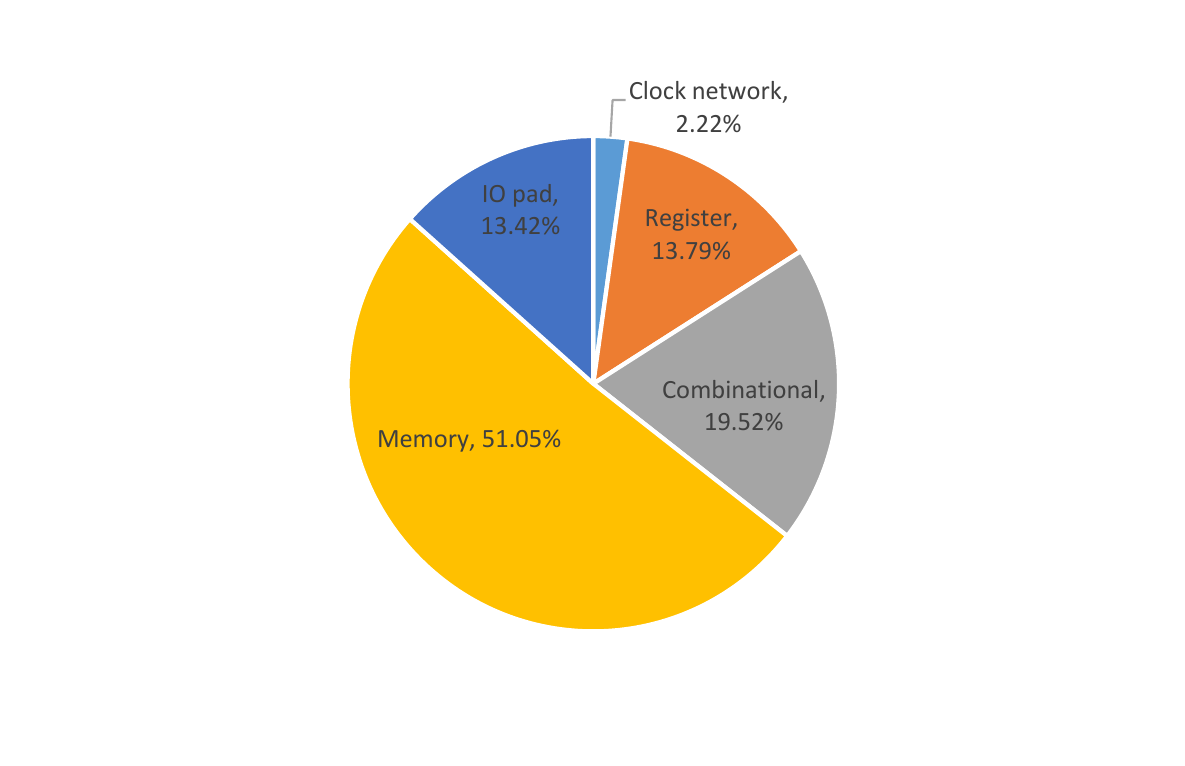}}
	\caption{The power breakdown of the chip.}
	\label{16-power}
\end{figure}
\subsection{Area and Power analysis}
For the presented chip, on-chip SRAM occupies 63.9\% of the area due to the large size of the unified buffer and weight buffer for layer fusion. For the logic area, the PE array occupies 42\% of the area due to 768 MAC. The accumulator occupies 28\% of the area due to 24-bits adders and few FIFO registers for partial sum accumulation. The controller occupies 21\% of the area due to a large number of multiplexers and wire routing between the buffer and the DLA core for layer fusion processing. 


Fig.~\ref{16-power} shows the power breakdown of the chip. Memory access accounts for 51\% of power due to the large amount of access for layer fusion. However, we only need to access the original feature map and its output feature maps once through I/O pads. Thus, I/O pads consume 13.4\% of the total power. The combinational logic consumes 19.5\% of power due to 768 PEs, the pipelined accumulator, and the processing of BN and ReLU6. The register logic consumes 13.7\% because most of the data are stored in on-chip memory, and thus needs few registers for pipeline hardware and accumulation. This also results in lower power consumption in the clock network, which accounts 2.2\% of total power. The core energy for 1280x720@30fps is 692.3mJ. 

When comparing with our prior design\cite{VWA} with the same PE numbers and model, the area overhead due to group fusion is the larger buffer. However, if the buffers for both designs are partitioned into the same sized banks, the energy consumption of these two designs will be comparable due to the same internal memory access amount and computation.

Above power consumption is only for the chip itself. The main contribution of this design is the low external memory traffic. The energy of the external DRAM access is reduced by 7.9X, from 2607mJ to 327.6mJ, as shown in Table \ref{tab:membw}, when compared to our prior design\cite{VWA} with the same PE numbers and model.

\subsection{Design Comparison}
Table~\ref{table:overview} shows comparisons with other designs, which is difficult due to different tasks and fusion design. The peak throughput of the proposed design can reach up to 460.8 GOPS. The area efficiency is 0.25 GOPS/KGE and 101.05 GOPS/mm$^2$. The area efficiency is better than most of the designs except \cite{MTK_fusion_isscc} due to its 7 nm process. 

The core power consumption is 692.3mW. The core power efficiency can reach up to 0.66 TOPS/W because of the simple and regular data flow. 
Our power efficiency is higher than \cite{eyeriss, eyerissv2, MTK_fusion_isscc} but lower than Envision\cite{envision}, SRNPU\cite{SRNPU_JETCAS}, and THINKER\cite{thinker_2018jssc} after normalized power efficiency. Envision\cite{envision} has higher power efficiency only when using lower precision hardware to compute 30\% - 60\% sparse network at lower supply voltage. These low power techniques can be applied to our design as well if needed. Thinker\cite{thinker_2018jssc} saves SRAM access power by using area hungry DFFs and thus has much higher area cost than our design.  However, both Envision and THINKER are nonfusion designs. Thus, their external DRAM access will be similar to our prior design with the same PE numbers when executing the same model, which is significantly higher than our design and will cancel out the benefits of the lower chip power consumption. 

For fusion-based designs, SRNPU\cite{SRNPU_JETCAS} designs a small dedicated network for super resolution. The total model size is around 130K, which can be stored on chip. This is not easily transferable to other networks. It has higher power efficiency than our design, but its area efficiency is much lower due to the large amount of small size cache accessing, more complex routing, and larger buffer area. Lin et al. \cite{MTK_fusion_isscc} uses mega scale on-chip buffer for fusion and has lower power efficiency than our design. Both designs use the direct layer fusion, which cannot fully exploit the benefit of layer fusion. Besides, they do not consider hardware utilization for fusion. 

In summary, the proposed fusion design has achieved a competitive performance for its core design compared to others. In addition, our design saves significant external DRAM energy and has higher area efficiency than others.

\section{Conclusion}
This paper proposes a deep learning chip with a low memory bandwidth for real-time HD object detection. We use group fusion to solve the high external memory traffic. The proposed resource-constrained network fusion and pruning, as well as the unified buffer design, enable this fusion. When compared to the previous naive fusion approach, this approach reduces the external memory traffic from 80.45 MB to 21.55 MB for 1920x960 input with a weight buffer size of 100 KB. This fusion also included hardware-oriented guidelines to maximize hardware utilization and reduce the bandwidth by 7.9 times to 585 MB/s for HD images with 72.12\% and 80.81\% mAP for the Pascal VOC 2007 and HD size datasets, respectively. The final chip implemented on a TSMC 40 nm CMOS process can execute real-time object detection at 1280x720@30FPS while reducing the external DRAM access energy from 2607mJ to 327.6mJ. The proposed method can also be easily integrated into other existing DLAs to improve energy consumption.





\bibliographystyle{IEEEtran}
\bibliography{IEEEabrv,thesis}

\begin{IEEEbiography}[{\includegraphics[width=1in,height=1.25in,clip,keepaspectratio]{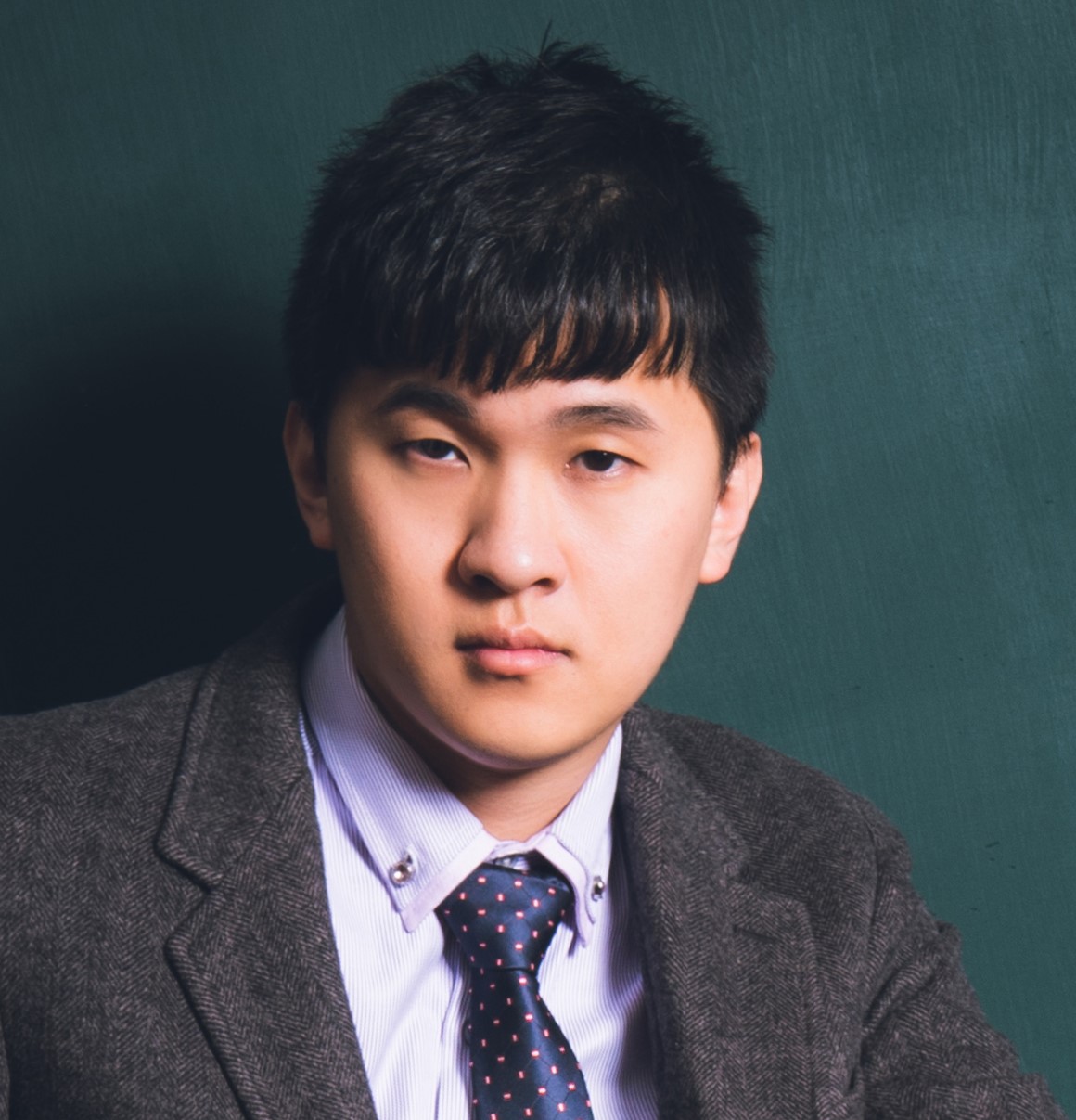}}]{Kuo Wei Chang}
recevied the Ph.D. degree from the National Chiao Tung University(NCTU), Hsinchu, Taiwan, R.O.C., in 2020. He is currently working in the Novatek, Hsinchu, Taiwan, R.O.C. His research interests include SoC design and deep learning.

\end{IEEEbiography}

\begin{IEEEbiography}[{\includegraphics[width=1in,height=1.25in,clip,keepaspectratio]{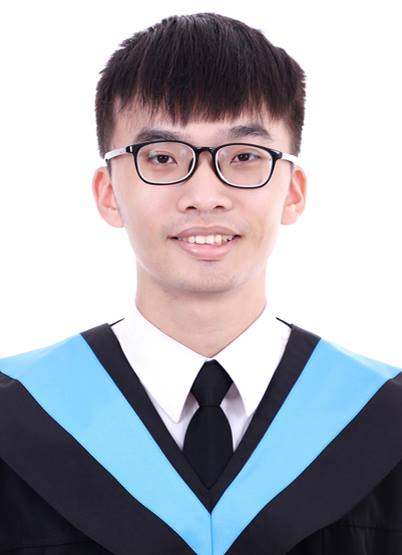}}]{Hsu Tung Shih}
recevied the B.S. and M.S. degree from the National Chiao Tung University(NCTU), Hsinchu, Taiwan, R.O.C., in 2018, and 2020. He is currently working in the Realtek, Hsinchu, Taiwan, R.O.C. His research interest is deep learning.

\end{IEEEbiography}

\begin{IEEEbiography}[{\includegraphics[width=1in,height=1.25in,clip,keepaspectratio]{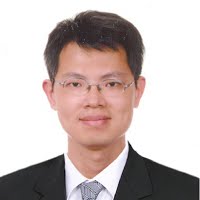}}]{Tian-Sheuan Chang}
	(S’93–M’06–SM’07)
	received the B.S., M.S., and Ph.D. degrees in electronic engineering from National Chiao-Tung University (NCTU), Hsinchu, Taiwan, in 1993, 1995, and 1999, respectively. 
	
	From 2000 to 2004, he was a Deputy Manager with Global Unichip Corporation, Hsinchu, Taiwan. In 2004, he joined the Department of Electronics Engineering, NCTU (as National Yang Ming Chiao Tung University (NYCU) in 2021), where he is currently a Professor. In 2009, he was a visiting scholar in IMEC, Belgium. His current research interests include system-on-a-chip design, VLSI signal processing, and computer architecture.
	
	Dr. Chang has received the Excellent Young Electrical Engineer from Chinese Institute of Electrical Engineering in 2007, and the Outstanding Young Scholar from Taiwan IC Design Society in 2010. He has been actively involved in many international conferences as an organizing committee or technical program committee member.
\end{IEEEbiography}

\begin{IEEEbiography}[{\includegraphics[width=1in,height=1.25in,clip,keepaspectratio]{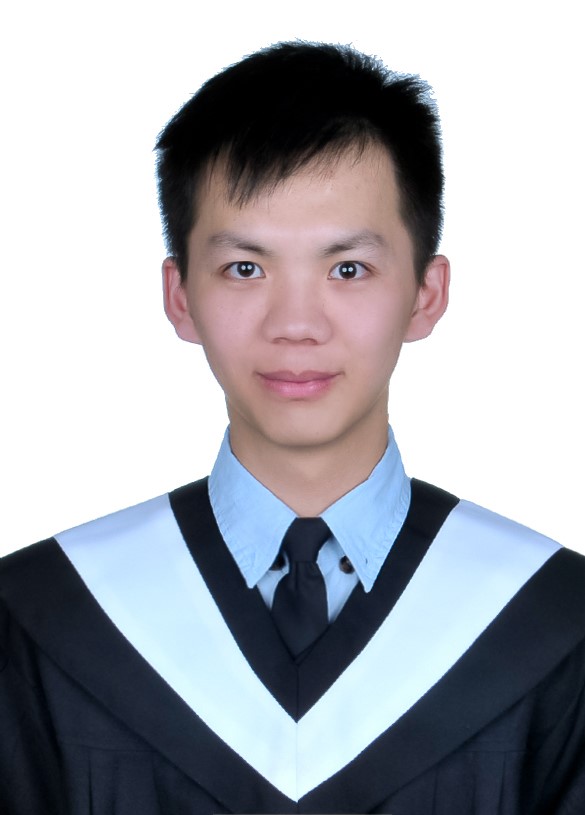}}]{Shang Hong Tsai}
received the B.S. degree in electrical engineering from the National Cheng Kung University, Tainan, Taiwan, R.O.C., and the M.S. degree in electronics engineering from National Chiao Tung University, Hsinchu, Taiwan R.O.C.. He is currently a principle engineer at Taiwan Semiconductor Research Institute (TSRI), Taiwan. His research interests include VLSI design and hardware implementation.

\end{IEEEbiography}

\begin{IEEEbiography}[{\includegraphics[width=1in,height=1.25in,clip,keepaspectratio]{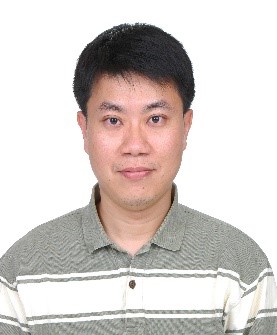}}]{Chih-Chyau Yang}
received the B.S. degree in electrical engineering from the National Cheng Kung University, Tainan, Taiwan, R.O.C. in 1996, and the M.S. degree in electronics engineering from National Chiao Tung University, Hsinchu, Taiwan R.O.C. in 1999. He is currently a principal engineer at Taiwan Semiconductor Research Institute (TSRI), Taiwan. His research interests include VLSI design, computer architecture, and platform-based SoC design methodologies.
\end{IEEEbiography}

\begin{IEEEbiography}[{\includegraphics[width=1in,height=1.25in,clip,keepaspectratio]{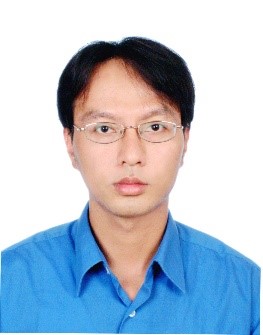}}]{Chien-Ming Wu}
obtained his B.S. and M.S degrees, both in electronic engineering, from National Yunlin University of Science and Technology, Taiwan, in 1997 and 1999, respectively. He received the Ph.D. degree from the Graduate School of Engineering Science and Technology at National Yunlin University of Science and Technology, Taiwan, in 2003. He is currently a researcher fellow and deputy division manager at Taiwan Semiconductor Research Institute (TSRI), Taiwan. His research interests include VLSI design in communication, coding theory, platform-based SoC design, and digital signal processing. 
\end{IEEEbiography}

\begin{IEEEbiography}[{\includegraphics[width=1in,height=1.25in,clip,keepaspectratio]{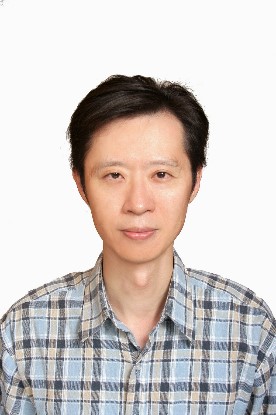}}]{Chun-Ming Huang}
received the B.S. degree in mathematical science from National Chengchi University, Taipei, Taiwan, R.O.C., in 1990, and the M.S. and Ph.D. degree, both in computer science, from the National Tsing-Hua University, Hsin-Chu, Taiwan, R.O.C., in 1992 and 2005, respectively. He is currently a researcher fellow and division manager at Taiwan Semiconductor Research Institute (TSRI), Taiwan. His research interests include VLSI design and testing, platform-based SOC design methodology, and multimedia communication. Dr. Huang is a member of Phi Tau Phi Scholastic Honor Society. 

\end{IEEEbiography}

\end{document}